\documentclass[aps,prb,twocolumn,superscriptaddress,oneside,floatfix,amsmath,showpacs,amssymb,footinbib]{revtex4-1}
\usepackage{graphicx}
\usepackage{dcolumn}
\usepackage{bm}
\usepackage{commath}
\begin{document}

\newcommand{\Hg}{HgBa$_2$CuO$_{4+\delta}$}
\newcommand{\Y}{YBa$_2$Cu$_3$O$_{6+\delta}$}
	
%%%%%%%%%%%%%%%%%%%%%%%%%%%% TITLE
	
\title{Synchrotron x-ray scattering study of charge-density-wave order in \Hg}

%%%%%%%%%%%%%%%%%%%%%%%%%%%% AUTHORS
	
	\author{W.~Tabis}
	\email{wojciech.tabis@lncmi.cnrs.fr}
	\affiliation{School of Physics and Astronomy, University of Minnesota, Minneapolis, Minnesota 55455, USA}
	\affiliation{AGH University of Science and Technology, Faculty of Physics and Applied Computer Science, 30-059 Krakow, Poland}
	\affiliation{Laboratoire National des Champs Magnetiques Intenses (CNRS, INSA, UJF, UPS), 31400 Toulouse, France}
	
	\author {B. Yu}
	\affiliation{School of Physics and Astronomy, University of Minnesota, Minneapolis, Minnesota 55455, USA}
	
	\author {I. Bialo}
	\affiliation{AGH University of Science and Technology, Faculty of Physics and Applied Computer Science, 30-059 Krakow, Poland}
	
	\author {M. Bluschke}
	\affiliation{Helmholtz-Zentrum Berlin fur Materialien und Energie, D-12489 Berlin, Germany}
	\affiliation{Max Planck Institute for Solid State Research, D-70569 Stuttgart, Germany}
	
	\author {T. Kolodziej}
	\affiliation{AGH University of Science and Technology, Faculty of Physics and Applied Computer Science, 30-059 Krakow, Poland}
	\affiliation{Advanced Photon Source, Argonne National Laboratory, Lemont, IL 60439, USA}
	
	\author {A. Kozlowski}
	\affiliation{AGH University of Science and Technology, Faculty of Physics and Applied Computer Science, 30-059 Krakow, Poland}
	
	\author {E. Blackburn} 
	\affiliation{School of Physics and Astronomy, University of Birmingham, Birmingham, B15 2TT, UK}
	
	\author{K. Sen}
	\affiliation{Institut f\"ur Festk\"orperphysik, Karlsruher Institut f\"ur Technologie, 76021 Karlsruhe, Germany}
	
	\author{E. M. Forgan}
	\affiliation{School of Physics and Astronomy, University of Birmingham, Birmingham, B15 2TT, UK}
	
	\author{M. v. Zimmermann}
	\affiliation{Deutsches Elektronen-Synchrotron DESY, 22603 Hamburg, Germany}

	\author {Y. Tang}
	\affiliation{School of Physics and Astronomy, University of Minnesota, Minneapolis, Minnesota 55455, USA}
	
	\author {E. Weschke}
	\affiliation{Helmholtz-Zentrum Berlin fur Materialien und Energie, D-12489 Berlin, Germany}
	
	\author {B. Vignolle}
	\affiliation{Laboratoire National des Champs Magnetiques Intenses (CNRS, INSA, UJF, UPS), 31400 Toulouse, France}
	
	\author {M. Hepting}
	\author {H. Gretarsson}
	\affiliation{Max Planck Institute for Solid State Research, D-70569 Stuttgart, Germany}
	
	\author{R.~Sutarto}
	\affiliation{Canadian Light Source, Saskatoon, Saskatchewan S7N 2V3, Canada}
	
	\author{F. He}
	\affiliation{Canadian Light Source, Saskatoon, Saskatchewan S7N 2V3, Canada}
	
	\author{M. Le Tacon}
	\affiliation{Institut f\"ur Festk\"orperphysik, Karlsruher Institut f\"ur Technologie, 76021 Karlsruhe, Germany}
	
	\author{N. Bari\v si\' c}
	\affiliation{School of Physics and Astronomy, University of Minnesota, Minneapolis, Minnesota 55455, USA}
	\affiliation{Institute of Solid State Physics, TU Wien, 1040 Vienna, Austria}
	
	\author {G. Yu}
	\affiliation{School of Physics and Astronomy, University of Minnesota, Minneapolis, Minnesota 55455, USA}
	
	\author{M.~Greven}
	\email{greven@umn.edu}
	\affiliation{School of Physics and Astronomy, University of Minnesota, Minneapolis, Minnesota 55455, USA}
	
	\date{\today}
	
	%%%%%%%%%%%%%%%%%%        ABSTRACT         %%%%%%%%%%%%%%
	
	\begin{abstract}
	We present a detailed synchrotron x-ray scattering study of the charge-density-wave (CDW) order in simple tetragonal \Hg~(Hg1201). 
	Resonant soft x-ray scattering measurements reveal that short-range order appears at a temperature that is distinctly lower than the pseudogap temperature and in excellent agreement with a prior transient reflectivity result. 
	Despite considerable structural differences between Hg1201 and \Y, the CDW correlations exhibit similar doping dependences, and we demonstrate a universal relationship between the CDW wave vector and the size of the reconstructed Fermi pocket observed in quantum oscillation experiments. 
	The CDW correlations in Hg1201 vanish already below optimal doping, once the correlation length is comparable to the CDW modulation period, and they appear to be limited by the disorder potential from unit cells hosting two interstitial oxygen atoms.
	A complementary hard x-ray diffraction measurement, performed on an underdoped Hg1201 sample in magnetic fields along the crystallographic $c$ axis of up to 16 T, provides information about the form factor of the CDW order. 
	As expected from the single-CuO$_2$-layer structure of Hg1201, the CDW correlations vanish at half-integer values of $L$ and appear to be peaked at integer~$L$. 
	We conclude that the atomic displacements associated with the short-range CDW order are mainly planar, within the CuO$_2$ layers.

	\end{abstract}
	%\pacs{61.05.C−, 74.25.Dw, 74.72.Jt, 74.62.Dh}
	% X-ray diffraction and scattering, 61.05.C−
	% Superconductivity, Phase diagrams, 74.25.Dw
	% Superconductivity, Other cuprates, including Tl and Hg-based cuprates, 74.72.Jt
	% Superconductivity, Effects of crystal defects, doping and substitution 74.62.Dh
	\maketitle

%%%%%%%%%%%%%           INTRODUCTION          %%%%%%%%%%%%%%%%%%%%
\section{\label{sec:level1}INTRODUCTION}

\begin{figure}
	\includegraphics[width=1\linewidth,angle=0,clip]{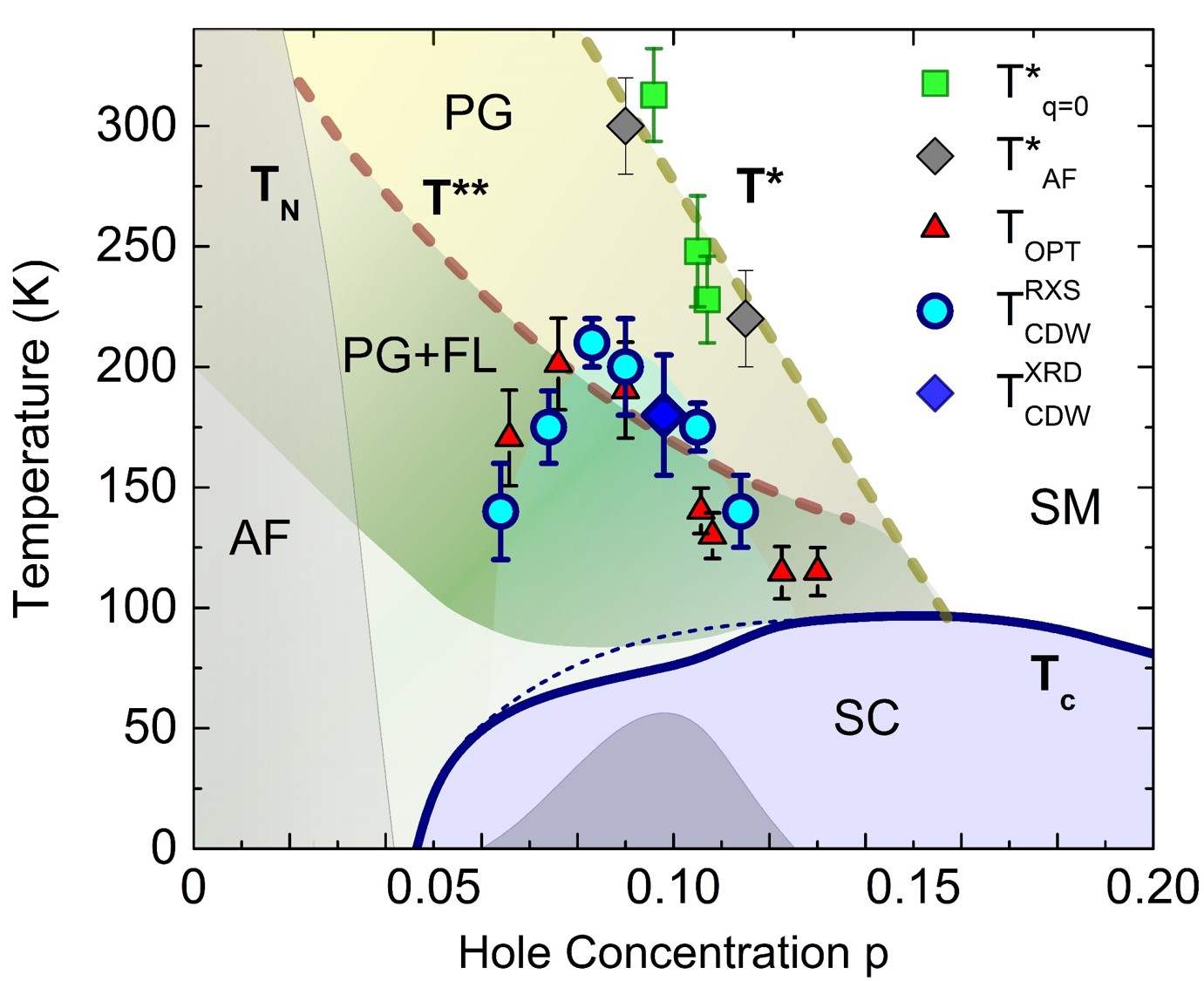}
	\caption{ (color online) Phase diagram of Hg1201 ($p>0.04$), extended to $p=0$ based on data for \Y~(YBCO). \cite{Barisic13b,Coneri10} Solid blue line: doping dependence of the superconducting (SC) transition temperature $T_c (p)$. \cite{Yamamoto00} Dark-grey region: deviation (not to scale) of $T_c$ from estimated parabolic doping dependence (dashed blue line). $T$*: pseudogap (PG) temperature, estimated from deviation of $T$-linear planar resistivity in the strange-metal (SM) state. \cite{Barisic13b,Barisic15} $T$**: temperature below which Fermi-liquid (FL) behavior is observed in the PG state. \cite{Barisic13b,Mirzaei13,Chan14,Barisic15} $T_{q=0}$ and $T_{AF}$: characteristic temperatures below which $q = 0$ magnetic order \cite{Li08,Li11b} and an enhancement of antiferromagnetic (AF) fluctuations \cite{Chan16a,Chan16b} are observed. $T_{CDW}$: Onset of short-range CDW order estimated from RXS (blue circles; $p=0.09$ results from Ref. \cite{Tabis14}) and XRD (blue diamond). $T_{opt}$: characteristic temperature observed in time-resolved optical reflectivity measurements. \cite{Hinton16}}
	\label{fig1}
\end{figure}

The discovery of superconductivity in the lamellar cuprates three decades ago triggered a tremendous amount of scientific activity, yet it has been a challenge to understand the strange-metal (SM) and pseudogap (PG) states from which the superconducting (SC) state evolves upon cooling (Fig.~\ref{fig1}). \cite{Keimer15} The extreme cases of zero and high hole-dopant concentrations are well understood: the undoped parent compounds  are antiferromagnetic (AF) Mott insulators,  whereas the highly-doped materials exhibit the characteristics of a conventional Fermi-liquid (FL) metal with a large Fermi surface (FS). 
The PG phenomenon is associated with myriad ordering tendencies, e.g., intra-unit-cell ($q=0$) magnetism that preserves the translational symmetry of the crystal lattice \cite{Fauque06,Li08} and a significant enhancement of dynamic, short-range  AF correlations. \cite{Chan16a,Chan16b} Evidence for charge order (in the form of charge-spin ``stripes") was first reported for the La-based cuprates. \cite{Tranquada95} Subsequent STM, \cite{Hoffman02,Howald03} NMR \cite{Wu11} and x-ray scattering work \cite{Ghiringhelli12,Chang12} found that charge-density-wave (CDW) correlations are a universal characteristic of the underdoped cuprates. 

Among the central goals in current cuprate research is to firmly establish the universality and hierarchy of ordering tendencies in the PG state and to shed light on the connection between CDW and SC order. 
A second goal is to achieve a consistent understanding of CDW-related phenomena observed with distinct probes.
A third goal is to establish the connection between CDW order and charge transport. 
The CDW correlations develop in the part of the phase diagram where normal-state FL behavior has been observed. \cite{Barisic13b,Mirzaei13,Chan14,Barisic15} 
Moreover, CDW correlations are thought to be responsible for the FS reconstruction implied by low-temperature transport experiments in high magnetic fields. \cite{Doiron-Leyraud07,Bangura08,LeBoeuf07,Doiron-Leyraud13,Barisic13a,Chan16c} 
Finally, it is essential to understand the role of disorder, which might both stabilize otherwise fluctuating CDW order and limit the spatial extent of CDW domains. \cite{Caplan15} 

Here we tackle these questions through a systematic Cu $L-$edge resonant x-ray scattering (RXS) and nonresonant hard x-ray diffraction (XRD) study of the doping and temperature dependence of the CDW correlations in \Hg~(Hg1201). Hg1201 is ideal for experimental study due to its simple tetragonal crystal structure, with only one CuO$_2$ layer per primitive cell, no intervening CuO chains, and an optimal $T_c$ of nearly 100 K, the highest of all such single-layer compounds. All cuprates exhibit inherent inhomogeneity and disorder, \cite{Phillips03,Eisaki04} and Hg1201 is no exception in this regard. \cite{Bobroff97, Rybicki09}
In particular, the interstitial oxygen atoms in underdoped samples are likely randomly distributed in the HgO$_{\delta}$ layer ($\delta \approx 0.18$ at optimal doping), \cite{Abakumov98,Antipov98,Putilin00} yet the quasiparticle mean-free path and the screening of the Coulomb repulsion have been estimated to be rather large, which may contribute to the high $T_c$ of Hg1201. \cite{Chen11}
This is supported by the observation of a tiny residual resistivity, \cite{Barisic13b} of Shubnikov-de-Haas oscillations, \cite{Barisic13a,Chan16c} and of Kohler scaling of the magnetoresistance in the PG state. \cite{Chan14} Samples of comparable quality than those investigated in the present work have enabled state-of-the-art electronic Raman scattering, optical spectroscopy, and hard x-ray experiments. \cite{Homes04,Lu05,Li12,Li13,Heumen07,Heumen09,Wang14a,Hinton16,Mirzaei13,Cliento14}
Such Hg1201 samples feature a small density of vortex pinning centers, \cite{Barisic08} which has enabled the observation of a magnetic vortex lattice. \cite{Li11a}

This paper is organized as follows: In Section II, we provide sample information and further describe the experimental techniques. In Section III, we present the experimental results, which then are further discussed in Section IV. The conclusions follow in Section V. 
Additional information on the calculations and simulations presented in this article are detailed in Appendices A, B and C. The temperature-doping phase diagram is shown
in Fig. \ref{fig1} and sample information is summarized in Table I.

\begin{table*}
	\begin{tabular}{ p{1.5cm}| p{1.5cm} || p{2.5cm}  p{2.5cm}  p{2.5cm}  p{4cm}}
		%{|c|c|c c c c|}
		\hline \hline 
		\hspace{0.05cm} $T_c$ (K) & \hspace{0.05cm} $p$ & \hspace{0.4cm} $\xi/a$ & $q_{CDW}$ (r.l.u.) & $T_{CDW}$ (K) & Comments \\
		\hline \hline
		\hspace{0.05cm} 55 & \hspace{0.05cm} 0.064 & \hspace{0.4cm} 4.6(5) & 0.292(7) & 140(20) & very weak CDW\\ 
		\hline 
		\hspace{0.05cm} 65 & \hspace{0.05cm} 0.074 & \hspace{0.4cm} 5.8(5) & 0.297(5) & 175(15) & \\ 
		\hline 
		\hspace{0.05cm} 69 & \hspace{0.05cm} 0.083 & \hspace{0.4cm} 8.0(7) & 0.281(8) & 210(10) & \\ 
		\hline 
		\hspace{0.05cm} 72 & \hspace{0.05cm} 0.09 & \hspace{0.4cm} 6.4(6) & 0.276(6) & 200(20) & from ref. \cite{Tabis14}\\ 
		\hline
		\hspace{0.05cm} 74 & \hspace{0.05cm}  0.094 & \hspace{0.4cm} 6.0(5) & 0.289(7) & & \\
		\hline 
		\hspace{0.05cm} 75 & \hspace{0.05cm} 0.098 & \hspace{0.4cm} 7.0(7) & 0.278(5) &180(25) & XRD in magnetic field \\
		\hline 
		\hspace{0.05cm} 76 & \hspace{0.05cm} 0.10  & \hspace{0.4cm} 5.2(4) & 0.270(5) & & \\ 
		\hline 
		\hspace{0.05cm} 79 & \hspace{0.05cm} 0.105 & \hspace{0.4cm}  4.6(3) & 0.270(5) & 175(10) &  \\ 
		\hline 
		\hspace{0.05cm} 88 & \hspace{0.05cm} 0.115 & \hspace{0.4cm} 4.8(7) & 0.260(6) & 140(50) & \\ 
		\hline 
		\hspace{0.05cm} 94 & \hspace{0.05cm} 0.126 & \hspace{0.4cm} - & - & - & no CDW\\ 
		\hline \hline
	\end{tabular} 
	\caption{Hg1201 sample information. The superconducting critical temperature was determined from the onset of the diamagnetic response in magnetic susceptibility measurements. The hole concentration was estimated from the $T_c$ vs. $p$ relationship established in ref. \cite{Yamamoto00} The CDW correlation length $\xi/a = (1/(\pi\cdot {\rm FWHM}))$, was determined near $T_c$, from the full-width-at-half-maximum (FWHM) of Gaussian fits to the data, as described in the text. Similarly, the wave vector was obtained from the center of the peak. $T_{CDW}$ was estimated as described in the text.}
	\label{tab:samples}   
\end{table*}

\section{\label{sec:level1} EXPERIMENTAL METHODS}

\subsection{\label{sec:level2} Samples}
Hg1201 single crystals were grown by a previously reported flux method, \cite{Zhao06} subsequently annealed to achieve the desired oxygen (and hence hole) concentration, \cite{Barisic08} and then quenched to room temperature. The superconducting transition was determined from magnetic susceptibility measurements (magnetic field strength 5 Oe), and the critical temperature $T_c$ was defined as the onset of the diamagnetic response. The hole concentration was estimated from the $T_{c}$ vs. $p$ relationship established for polycrystalline samples in ref. \cite{Yamamoto00} Sample information is summarized in Table \ref{tab:samples}. The typical sample dimensions were 2 x 2 x 0.5 mm$^3$, with the short side along the crystallographic $c$ axis. The x-ray penetration length in Hg1201 is about 0.2 $\mu$m at the energy used in the soft x-ray experiment ($\hbar\omega\approx$ 932 eV). In order to ensure flat and clean surfaces, the samples were polished multiple times with sandpaper with increasingly finer grade, ranging from 1 $\mu$m to 0.05 $\mu$m. The x-ray diffraction measurements at 80 keV were performed in transmission geometry and thus no particular surface preparation was required.

\subsection{\label{sec:level2} Resonant x-ray scattering}
Resonant x-ray scattering (RXS) measurements at the Cu $L_{3}$ edge were performed at the UE46 beam line of the BESSY-II synchrotron in Berlin, Germany, and at the REIXS beam line \cite{Hawthorn11} of the Canadian Light Source. In order to determine the exact resonance energy, x-ray absorption measurements were performed in total-fluorescence-yield configuration. The incident x-ray beam was then tuned to the maximum of the fluorescence signal for each individual sample ($\hbar\omega\approx$ 932 eV). Since the structure of Hg1201 contains only one Cu site per formula unit, a fluorescence spectrum displays a single maximum.\cite{Tabis14} Momentum scans were performed by rotating the sample about the axis perpendicular to the scattering plane, and the detector angle was set to 2$\theta = 160^{\circ}$. 
We quote the scattering wave-vector ${\bf Q} = H{\bf a^*} + K{\bf b^*} + L {\bf c^*} = (H, K, L)$ in reciprocal lattice units, where $a^* = b^* = 1.62$ \AA$^{-1}$ and $c^* = 0.66$ \AA$^{-1}$ are the approximate room-temperature values. In the configuration used in the experiment, $K$ was set to zero, and $H$ was coupled to $L$ during the scans. 
The maximum of the CDW peak was observed at doping dependent $H$ values (also denoted as $q_{CDW}$) in the approximate range 0.26 to 0.29 r.l.u., near the largest experimentally accessible value of $L \approx 1.25$ r.l.u. Due to the large x-ray wavelength at the Cu $L$-edge, only the first Brillouin zone was accessible, and hence ${\bf Q}$ is equivalent to the reduced wave vector. We denote the reduced two-dimensional wave vector as ${\bf q}$; CDW peaks are observed at the equivalent positions ${\bf q} = (q_{CDW},0)$ and $(0,q_{CDW})$.

\subsection{\label{sec:level2} Hard x-ray Diffraction}
Hard x-ray diffraction (XRD) measurements were carried out in transmission geometry on a $T_c = 75$ K sample using the 80 keV synchrotron radiation at the P07 beamline of the PETRA III storage ring at DESY, Hamburg, Germany. The 17-T horizontal cryomagnet provided by the University of Birmingham was installed on the triple-axis diffractometer.\cite{Holmes12} Following prior work on YBCO, \cite{Chang12} the magnetic field was used to enhance the CDW correlations and thus increase the sensitivity of the experiment at low temperatures. In order to achieve good mechanical stability and thermalization, the sample was mounted to a silicon wafer, then glued to a temperature-controlled aluminum plate within the cryomagnet vacuum, and thermally shielded by aluminized mylar foils glued to this plate. The sample temperature was controlled in the range $2 - 300$ K. The incident and scattered x rays passed through the kapton cryostat vacuum windows, which gave a maximum of $\pm10^{\circ}$ input and output angles relative to the field direction. The high energy of the x rays allowed us to access a significantly larger portion of the reciprocal space compared to the RXS experiment. In the XRD experiment, $H$-scans centered at the CDW peak position $q_{CDW}$ were performed across many Brillouin zones; the accessible values of $H$ (in the range of $\sim1-8$) were dependent on the $L$ values. The XRD experiment was performed in an applied magnetic field of 16 T, and the field component along the crystallographic $c$ axis (i.e., perpendicular to the CuO$_2$ planes) was dependent on ${\bf Q}$. For larger ${\bf Q}$ values ($L>2$), the angle between the $c$ axis of the crystal and the magnetic field direction was about $30^{\circ}$, which resulted in a maximum $c$-axis field of approximately 14 T. The magnetic field was applied at a temperature above $T_{c}$, and the sample was then field-cooled to base temperature. As a function of the magnetic field, minor changes in the position and angle of the sample holder were observed; these were corrected with horizontal and vertical motion stages situated under the cryostat rotation stage, and by realigning the sample on the (0~2~0) Bragg peak. Due to time constraints, no measurements of the CDW peak were performed in zero applied field.

\section{\label{sec:level1} RESULTS}

\subsection{\label{sec:level2} Doping dependence of the CDW order}

Building on prior work for $p \approx 0.09$ ($T_c \approx 72$ K), \cite{Tabis14} we studied crystals at eight doping levels in the range $0.064 \leq p \leq 0.126$ ($T_c \approx$ 55 - 94 K) via RXS. In addition, we measured one crystal via XRD at a doping level ($p \approx 0.098$, $T_c \approx 75$ K) at which the charge correlations are relatively strong. The onset temperature of the CDW order, $T_{CDW}$, was determined for a subset of these samples (Fig.~\ref{fig1}, Table \ref{tab:samples}). As demonstrated in Figs. 2 - 5, a CDW signal was observed at all doping levels, except for the most highly doped sample with $T_c \approx$ 94 K (Fig.~\ref{fig3}(b)). As demonstrated in Fig.~\ref{fig3}(c), the CDW order is observed at equivalent wave vectors (magnitude $q_{CDW}$) along [100] and [010], as expected, given the tetragonal structure of Hg1201.

\begin{figure}[t]
	\includegraphics[width=1\linewidth,angle=0,clip]{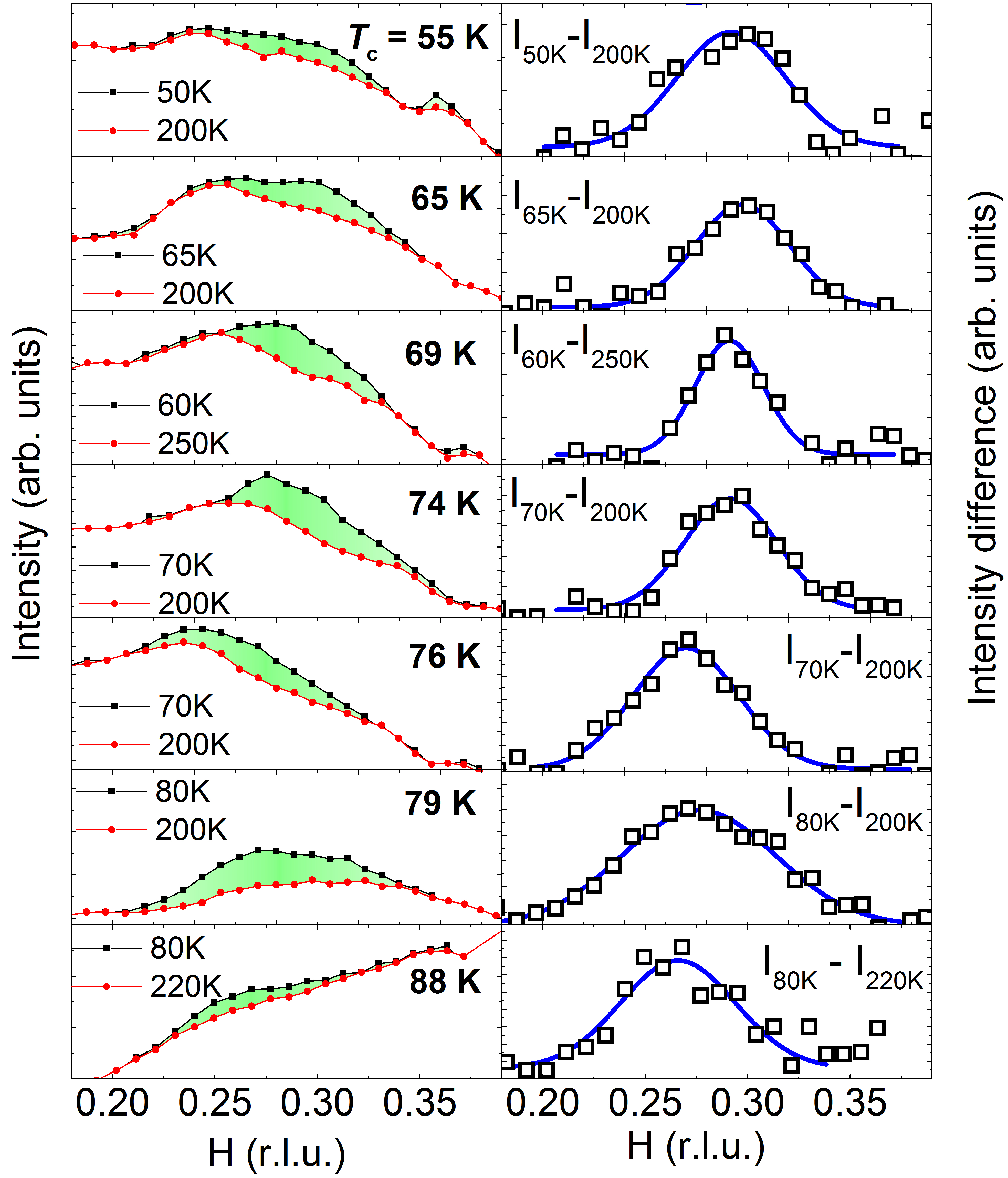}
	\caption{(color online) Momentum dependence of the CDW peak observed via RXS in seven underdoped samples along [100]. Left: raw spectra collected at indicated temperatures. Right: intensity difference of data in left panels. The high-temperature data constitute an estimate of the background scattering. Although the location of the CDW peak is not clear from the raw data, the background-subtracted scans clearly reveal CDW peaks. Blue lines: Gaussian fits to the data.}
	\label{fig2}
\end{figure}

\begin{figure}[h]
	\includegraphics[width=1\linewidth,angle=0,clip]{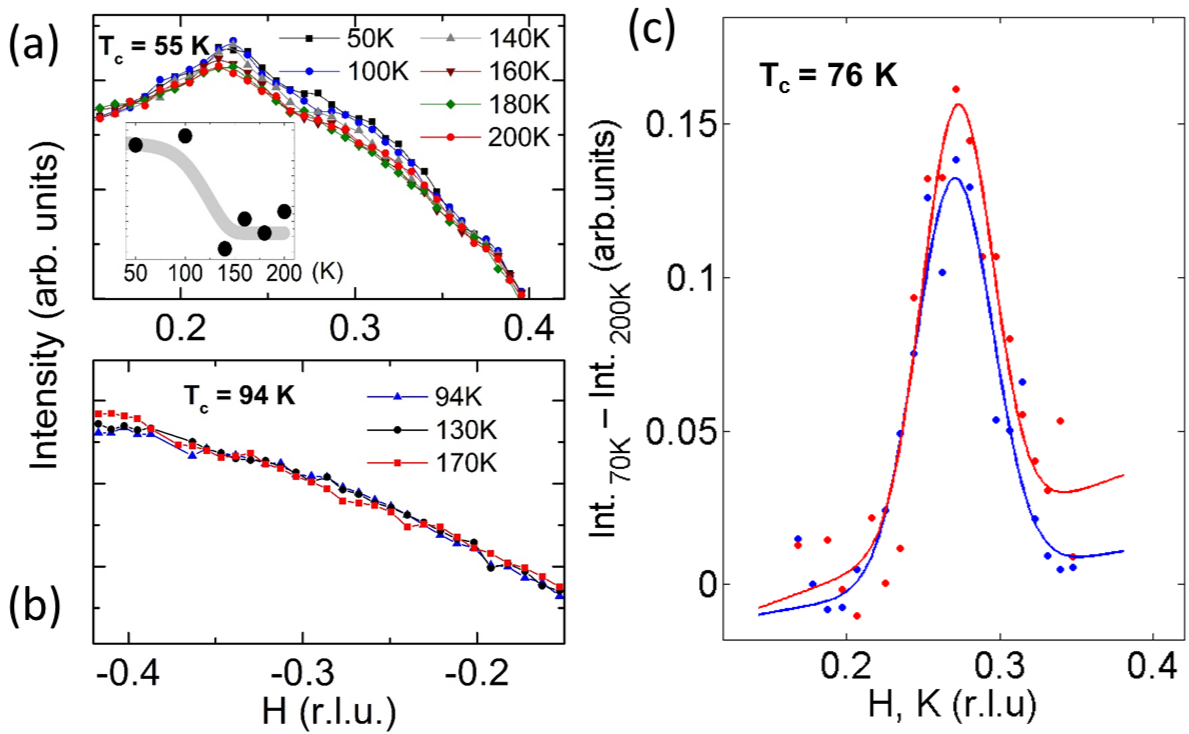}
	\caption{(color online) Momentum scans across the CDW wave vector at the Cu $L_3$ resonance at various temperatures for two Hg1201 samples with (a) $T_c$ = 55 K ($p$ = 0.064) and (b) $T_c$ = 94 K ($p$ = 0.126). Due to the very low intensity of the CDW peak in the first sample, systematic subtraction of high-temperature ``background'' from the low-temperature scans did not result in a reliable estimate of the temperature dependence. Instead, the onset temperature of the CDW order was determined from the temperature dependence of the integrated intensity (between $H = 0.2$ and 0.35 r.l.u.) after introducing an offset to match the background level at 200 K; the result is shown in the inset of (a). The onset temperature $T_{CDW}$ = 140(20)~K is the temperature at which the integrated intensity saturates. No intensity change with temperature is observed in the nearly optimally-doped sample ($T_c$ = 94 K), consistent with the absence of CDW order. (c) Background-subtracted momentum scan across $q_{CDW}$ for a sample with $T_c$ = 76 K ($p$ = 0.105) along the equivalent directions [100] and [010]. The solid lines are Gaussian fits (with linear background) to the  data, yielding equivalent peak positions, amplitudes and widths, within the experimental uncertainty.}
	\label{fig3}
\end{figure}

\begin{figure*}
	\includegraphics[width=1\linewidth,angle=0,clip]{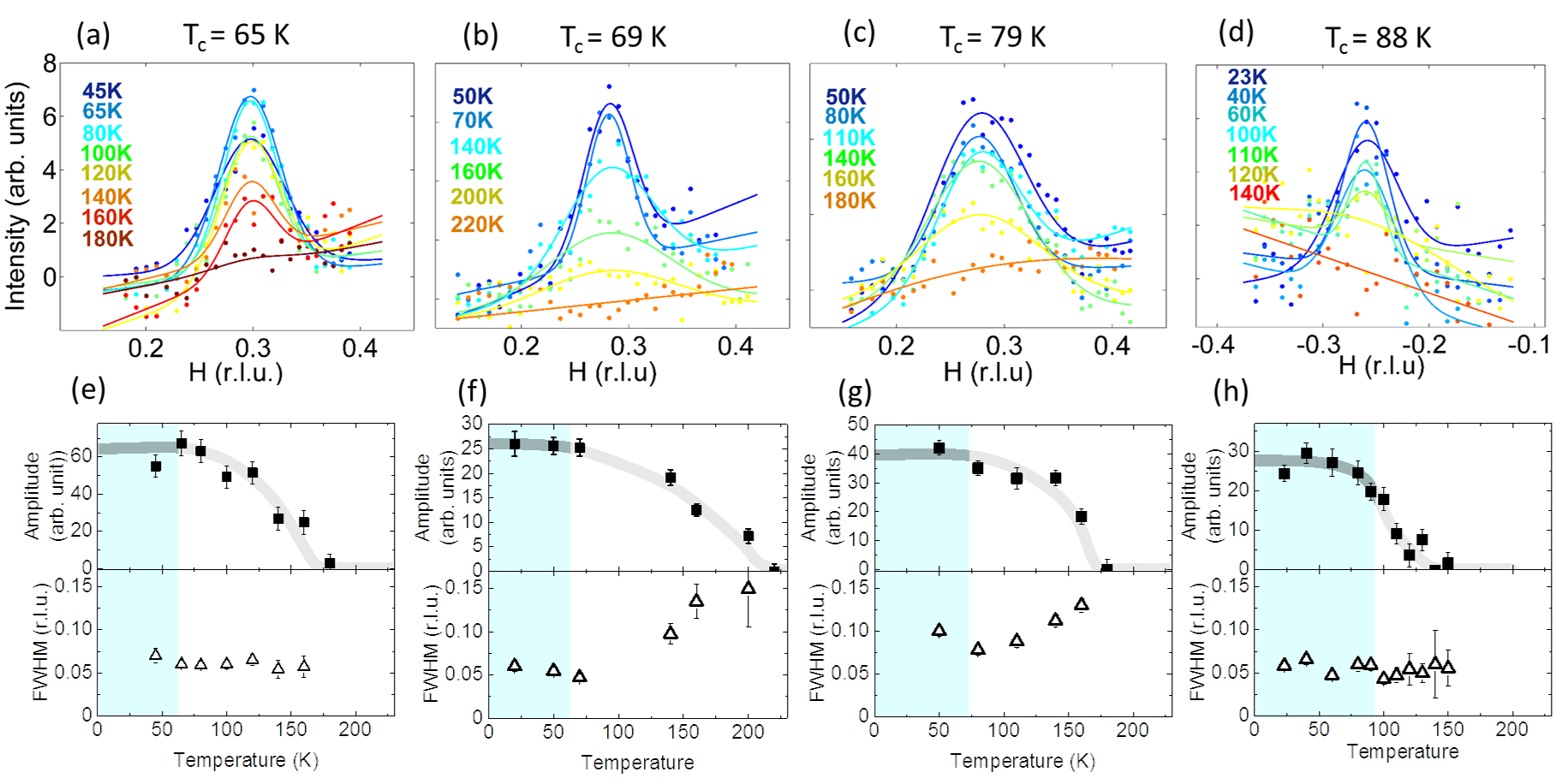}
	\caption{(color online) (a-d) Temperature dependence of the background-subtracted CDW peak for four Hg1201 samples observed by RXS. The peak intensity decreases above $T_c$ and saturates at the characteristic doping-dependent temperature $T_{CDW}$. In order to obtain the most accurate temperature dependence of the CDW peak parameters, high-temperature data were subtracted from the low-temperature scans (i.e., data at 200 K, 240 K, 200 K and 155 K were selected as reference background for the samples with $T_c$ = 65 K, 69 K, 79 K and 88 K, respectively). The solid lines are the result of subsequent Gaussian-function fits. (e-h) Temperature dependence of the peak amplitude and width (FWHM). Blue areas indicate the superconducting state for each sample. The thick gray lines are guides to the eye.}
	\label{RixTemp}
\end{figure*}

\begin{figure*}
	\includegraphics[width=1\linewidth,angle=0,clip]{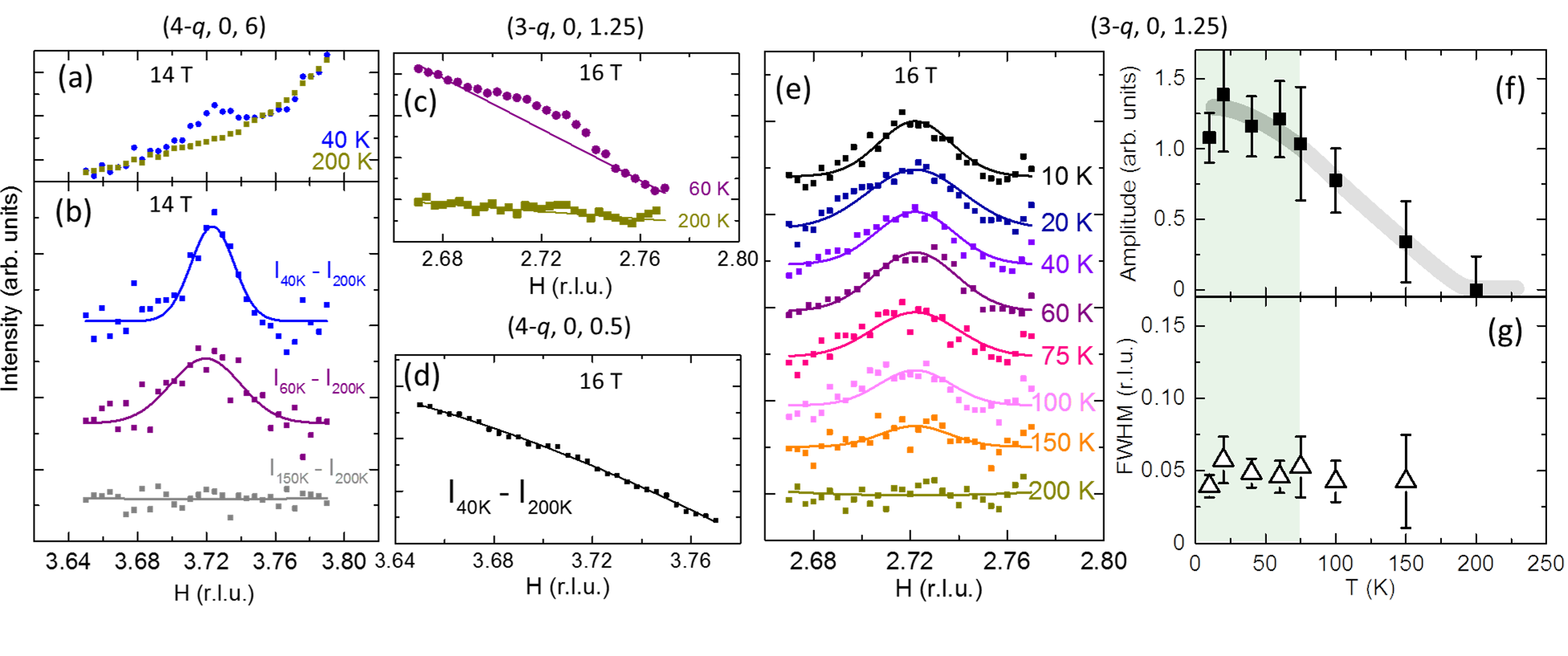}
	\caption{(color online) CDW peak in a Hg1201 sample, with $p = 0.098$ ($T_c = 75$ K), observed via 80 keV XRD  in various Brillouin zones, in an effective 14 T or 16 T $c-$axis magnetic field, as indicated. 
	The high energy of the hard x rays allowed access a large portion of momentum space. The corresponding Brillouin zone is marked above each panel. 
	(a) Momentum scans across the CDW wave vector at 40 K and 200 K. 
	(b) Intensity difference of the scans collected at the indicated temperatures. The solid lines are the result of subsequent fits to a Gaussian function. 
	(c) Two representative momentum scans across the CDW wave vector at 60 K and 200 K. The solid lines represent the estimated linear background. 
	(d) Corresponding intensity difference of the 40 K and 200 K data. These temperatures lie well below and above $T_{CDW}$, respectively. No clear evidence of CDW intensity is seen at half-integer $L$. 
	(e) Linear-background-subtracted data at the indicated temperatures. The solid lines are the results of fits to a Gaussian function. 
	(f-g) Temperature dependence of the CDW peak amplitude and width (FWHM) from the fit in (e). Green area indicates the temperature region of the superconducting state, in the absence of the magnetic field. Thick gray line in (f) is a guide to the eye. The data in (a-c) and (e) are offset for clarity.}
	\label{DiffAll}
\end{figure*}

Figure~\ref{SumAll} compares the CDW correlation length $\xi$ and peak position $q_{CDW}$ for Hg1201 (obtained from Gaussian fits; Figs. \ref{fig2}, \ref{DiffAll} and Ref.\cite{Tabis14}) with results for YBCO. \cite{Blanco-Canosa14} The ``SC domes" $T_c(p)$ of Hg1201 and YBCO differ somewhat,  \cite{Yamamoto00,Li08} and $p$ should be viewed as an effective parameter, \cite{Jurkutat14,Rybicki16} yet in both cases the maximum correlation length ($\xi_{max}$) corresponds to the center of the $T_c(p)$ plateau, where $T_{CDW}$ and the deviation from the interpolated parabolic $T_c(p)$ dependence are the largest (Fig.~\ref{fig1}). The suppression of $T_c$ appears to be material specific and stronger in YBCO than in Hg1201. \cite{Liang06,Yamamoto00} Whereas for Hg1201 the intensity of the CDW peak saturates in the SC state (Figs. \ref{RixTemp}, \ref{DiffAll}(f) and Ref.\cite{Tabis14}), for YBCO it decreases below $T_c$ (in the absence of an applied magnetic field). \cite{Blanco-Canosa14} For Hg1201, $\xi_{max}/a \approx 8$, which is about 40\% of $\xi_{max}/a \approx 20$ for YBCO. \cite{Blanco-Canosa14}  
Although $q_{CDW}$ is smaller than for YBCO, the respective rates of decrease with doping, $-0.6\pm0.1$ and $-0.49\pm0.05$ r.l.u./[holes/planar Cu], are the same within error (Fig.~\ref{SumAll}(b)). The different magnitudes reflect differences in the FS shapes of the two compounds. 

RXS is sensitive to the modification of the Cu $3d$ (valence) electronic states on nominally equivalent resonating ions. Whereas this technique is very sensitive to small CDW amplitudes, only a fraction of the first Brillouin zone can be reached due to the large wavelength of the x rays. As a result, RXS only allows the determination of the in-plane component of the CDW wave vector, i.e., the component parallel to the CuO$_2$ planes. On the other hand, (hard) XRD is sensitive to the entire charge cloud residing on an ion and probes atomic displacements associated with the charge order. The relatively short wavelength of hard x rays permits access to a large volume of reciprocal space. As shown in Figs.~\ref{DiffAll} and~\ref{SumAll}, the in-plane component of the CDW peak determined from XRD is consistent with the RXS results. The XRD measurement revealed a broad CDW signal along [001] that vanishes at half-integer $L$ values (Fig.~\ref{DiffAll}(d)) and appears to be strongest at the integer values of $L$ (Fig.~\ref{DiffAll}(a-b)).

\subsection{\label{sec:level2} Temperature dependence of the CDW order} 

Figure \ref{RixTemp} displays the temperature dependence of the CDW peak for four of our samples ($T_c$ = 65 K, 69 K, 79 K, and 88 K) measured via RXS. The corresponding temperature dependence for the $T_c$ = 75 K sample studied by XRD is shown in Fig. \ref{DiffAll}(e-f). Measurements were typically carried out at temperatures that ranged from 12 K to 250-300 K. The intensity variation was carefully analyzed in order to estimate the onset temperature $T_{CDW}$. 
For the RXS results, high-temperature data were used as reference ``background" and subtracted from each low-temperature scan. 
We checked that the intensity at q$_{CDW}$ does not vary further at higher temperatures. 
For the sample with the lowest doping level ($T_c = 55$ K), the CDW signal was too small to allow this analysis. Instead, as shown in Fig. \ref{fig3}(a), the intensity in the range $H=0.2-0.35$ r.l.u. was integrated at each temperature and the temperature at which the integrated intensity saturates upon heating was identified as $T_{CDW}$. 
Figure \ref{fig3}(b) demonstrates that no intensity variation was observed in the nearly optimally-doped sample ($T_c=94$ K).
In order to obtain the temperature dependence in the XRD measurement, a linear background was subtracted from the scans, as shown in Fig. \ref{DiffAll}(c).   
The background-subtracted peaks were fit to a Gaussian function (Fig. \ref{DiffAll}(e)), and the amplitude and FWHM was extracted. 
In overall agreement with the previous result for $T_c = 72$ K in ref.\cite{Tabis14}, the intensity around q$_{CDW}$ increases smoothly below $T_{CDW}$ and saturates below $T_c$.  

%%%%%%%%%%%%%%%         DISCUSSION        %%%%%%%%%%%%
	\section{\label{sec:level1} DISCUSSION}

\begin{figure}
	\includegraphics[width=1\linewidth,angle=0,clip]{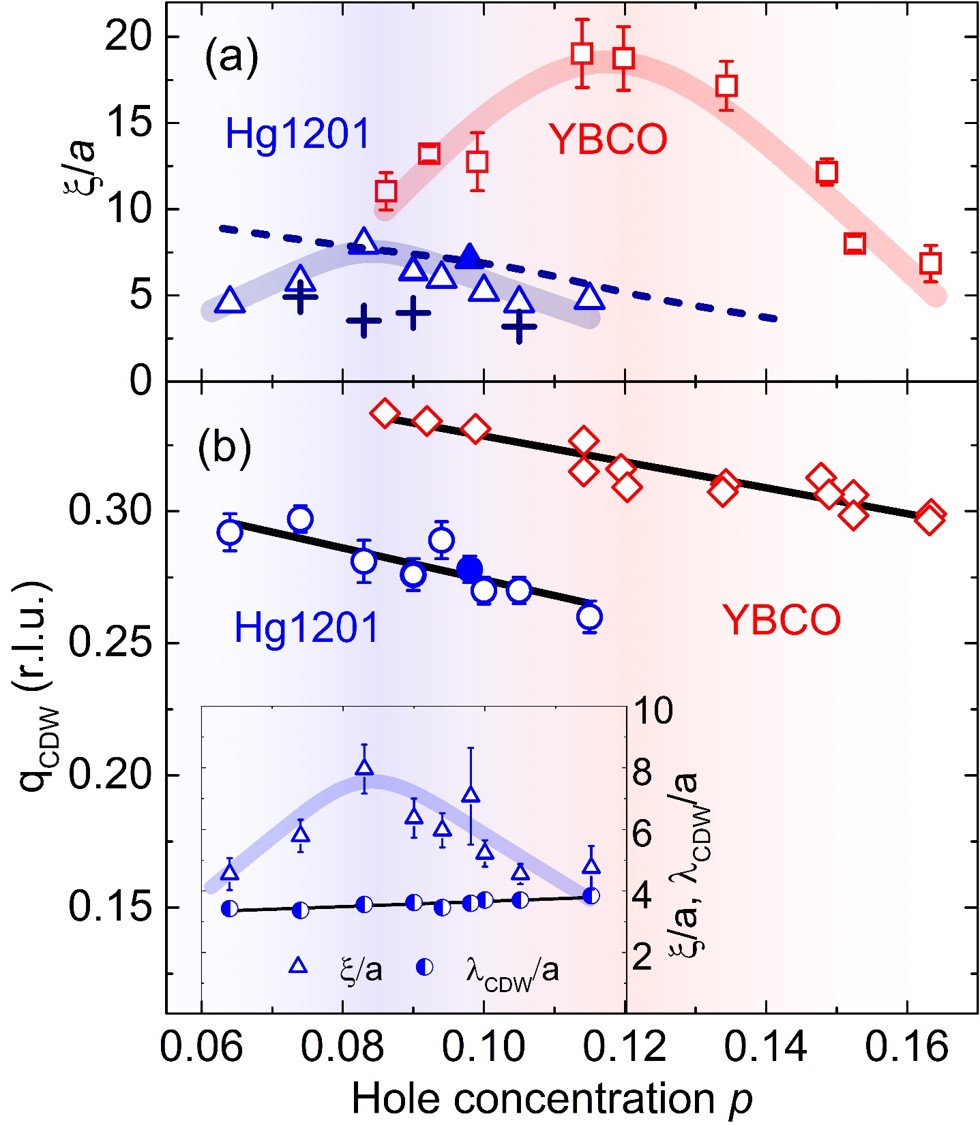}
	\caption{ (color online) (a) CDW correlation length (in units of the planar lattice parameter $a$) near $T_c$ for Hg1201. Open and filled blue triangles indicate RXS and XRD (at 14 T) results, respectively; the $p \approx 0.09$ result is from ref.\cite{Tabis14}. For comparison, the data for YBCO from ref. \cite{Blanco-Canosa14} are shown as red squares. For Hg1201, $\xi/a$ estimated about 20 K below $T_{CDW}$ is shown as well (blue crosses). Blue dashed line: simulated characteristic size of CDW domains, as discussed in the text. (b) $q_{CDW}$ for Hg1201 (blue circles) and YBCO (red diamonds). Black lines are linear fits (see text). Inset compares the correlation length of the CDW order in Hg1201 (obtained near $T_c$) with the corresponding wavelength $\lambda_{CDW}$, expressed in lattice units. The red and blue bands for $\xi(p)/a$ are guides to the eye.}
	\label{SumAll}
\end{figure}

There remain numerous open questions regarding the hierarchy of ordering tendencies in the PG state, the correlation between CDW signatures observed with different probes, the connection among CDW order, SC order and charge transport, the connection between the real-space charge modulations and the FS topology, and the role of disorder. 
The short-range CDW order appears well below $T_{AF}$ ($T_{AF} \approx T^*$), the temperature below which a distinct enhancement of AF fluctuations is observed, \cite{Chan16a,Chan16b} consistent with theoretical proposals that AF correlations drive the charge correlations \cite{Metlitski10,Efetov13,Wang14,Allais14,Atkinson15,Wang15} and with the notion \cite{Tabis14,Barisic15,Badoux16} that the PG and CDW formation are distinct phenomena.  
In Hg1201, the CDW correlations appear at or below $T^{**}$, the temperature below which FL charge transport is observed \cite{Barisic13b,Mirzaei13,Chan14} and below which the Hall constant is nearly independent of temperature, and thus the PG and associated FS is likely fully formed; \cite{Barisic15} only at the doping level where CDW order is most robust do $T_{CDW}$ and $T^{**}$ coincide. 
On the other hand, $T_{CDW}$ coincides at all doping levels with $T_{OPT}$, the characteristic temperature from transient reflectivity measurements, \cite{Hinton16} which suggests that both probes detect the same correlations.
As seen in Fig. \ref{SumAll}, at $p\approx0.115$, $\xi$ obtained near $T_c$ becomes comparable to the CDW modulation period (this is also the case close to $T_{CDW}$). 
Thus, the CDW loses spatial coherence already below optimal doping. 
Interestingly, a qualitative change in the behavior of the quasiparticle recombination rate ($\tau_{qp}$) was observed at about the same doping level: \cite{Hinton16}
whereas the behavior of $\tau_{qp}$ near optimal doping (where $T_{OPT}$ is not much larger than $T_c$) was found to be consistent with the mean-field BCS theory of superconductivity, the data at lower doping are best understood upon considering composite SC and CDW fluctuations near $T_c$. 
We note that CDW order was reported for optimally-doped ($p \approx 0.16$) Hg1201, \cite{Campi15} yet the very high onset temperature and the values for $\xi$ and $q_{CDW}$ are inconsistent with those extrapolated to optimal doping based on our study (see also Appendix \ref{OPD}).

\subsection{\label{sec:level2} CDW order and the FS reconstruction in the cuprates} 

Transport measurements have demonstrated the existence of a single quantum oscillation (QO) frequency in Hg1201, and hence of a single electron pocket at the Fermi level. \cite{Barisic13a,Chan16c} This is in stark contrast to orthorhombic double-layer YBCO which, in addition to the main oscillation frequency at $\sim$ 550 T, features satellites at $\pm$~100 T \cite{Vignolle13} due to magnetic breakdown between the complex electron pockets that originate from a coupling between adjacent CuO$_2$ layers, \cite{Sebastian12} and additionally a slow frequency ($\sim$ 100 T), possibly due to small hole-like FS pockets. \cite{Doiron-Leyraud15} 
Hg1201 thus is a pivotal system to seek a connection between CDW order and FS topology.
In ref.~\cite{Tabis14}, it was demonstrated for Hg1201 ($p \approx 0.09$) that folding the underlying FS with the measured $q_{CDW}$, and with assumed biaxial CDW order, results in a small electron pocket whose size agrees with the QO frequency observed at the same doping level \cite{Barisic13a,Chan16c}.
Similar agreement was also found for YBCO, but considering only the bonding FS and main QO frequency. 
Our result for the doping dependence of $q_{CDW}$ in Hg1201, in combination with a tight-binding FS calculation, allows us to simulate the doping dependence of the electron pocket and to compare with the QO data. \cite{Barisic13a,Chan16c}

\begin{figure}
	\includegraphics[width=1\linewidth,angle=0,clip]{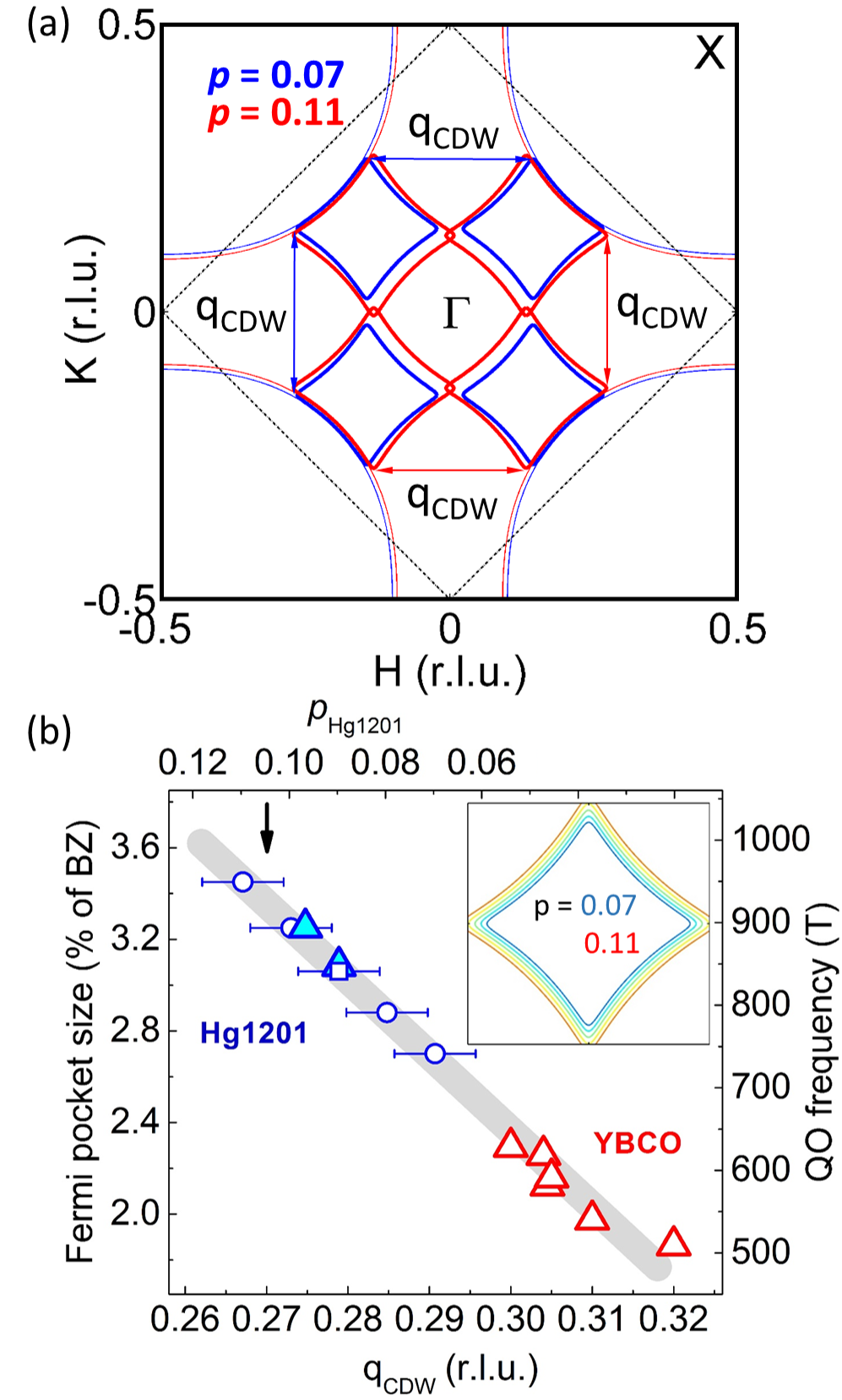}
	\caption{ (color online) (a) Noninteracting tight-binding FS of Hg1201 at $p=0.07$ (light blue) and $p=0.11$ (light red) with hopping parameters from Ref. \cite{Das12}. Thick lines: reconstructed electron pocket; double-arrows: $q_{CDW}$ (see main text). (b) Pocket size, as a fraction of the Brillouin zone, and associated quantum-oscillation frequency ($F$) for Hg1201 and YBCO as a function of $q_{CDW}$. Blue square (Hg1201) \cite{Tabis14,Barisic13a} and red triangles (YBCO) \cite{Vignolle13} represent the doping levels for which both $q_{CDW}$ and $F$ have been determined experimentally. $F$ is related to the Fermi pocket size $S$ via the Onsager relation, $F=\hbar S/2\pi e$. Blue circles: pocket size estimated from tight-binding calculation, with $q_{CDW}$ obtained from the linear fit in Fig.~\ref{SumAll}(b). Horizontal error bars: experimental uncertainty in $q_{CDW}$. Blue triangles: recent high-precision experimental result for $F$. \cite{Chan16c} Inset: doping dependence of the electron pocket in the reduced Brillouin zone (increments: $\Delta p = 0.01$). Predicted Lifshitz-type change in FS topology at $p\approx 0.105$ (black arrow) due to overlap of reconstructed pockets (see result for $p\approx 0.11$ in panel (a)).}
	\label{fig7}
\end{figure}

Figure~\ref{fig7}(a) shows the tight-binding FS of Hg1201. The electron pockets were obtained by solving a Hamiltonian linking the tight-binding FS, consistent with photoemission spectroscopy measurements, \cite{Vishik14} with the CDW wavevector values $q_{CDW}$ estimated from Fig.~\ref{SumAll}(b). Due to the PG in the antinodal regions, no additional hole-pockets are expected in the reconstructed state, consistent with experiment. \cite{Barisic13a,Chan16c} 
The resultant relationship between electron pocket size (or, equivalently, QO frequency) and q$_{CDW}$ (and $p$) is shown in Fig.~\ref{fig7}(b) (see Appendix \ref{FSC} for details). 
The high-precision QO data are in very good agreement with this prediction. \cite{Chan16c} 

Our measurements for tetragonal Hg1201 were conducted either in the absence of an applied magnetic field (RXS), or in relatively low magnetic fields compared to the upper critical field (XRD), and the data do not allow us to discern whether the short-range CDW order is unidirectional \cite{Comin15} or bidirectional \cite{Forgan15}. Nevertheless, the discussed FS reconstruction scenario points to a bidirectional character.
x-ray scattering work revealed that orthorhombic YBCO ($T_c = 67$ K) exhibits three-dimensional long-range CDW order in fields above about 18 T, \cite{Gerber15} consistent with prior NMR evidence for a field-induced phase transition. \cite{Wu11} 
This field-induced order is unidirectional and observed only along [010]. \cite{Gerber15,Chang16} 
NMR work suggests that the three-dimensional order does not simply evolve from the short-range two-dimensional  correlations, but rather that the latter coexist with the former. \cite{Wu15} 
Fermi-surface reconstruction does not necessarily require long-range CDW correlations. \cite{Chan16c}
Furthermore, Fermi-surface reconstruction by unidirectional (criss-crossed in a bilayer system) order would result in multiple QO frequencies, \cite{Maharaj16} which are not observed in Hg1201. \cite{Chan16c} 
Although multiple frequencies are observed in YBCO, \cite{Doiron-Leyraud07} this is not inconsistent with Fermi-surface reconstruction due to bidirectional CDW order. \cite{Allais14,Briffa16} 

Upon treating the hole concentration as an implicit parameter, a comparison between Hg1201 and YBCO can be made regarding the relationship between $q_{CDW}$ (measured in the absence of an applied magnetic field) and $F$ (estimated in the case of YBCO from the central QO frequency): Fig.~\ref{fig7}(b) demonstrates that these two observables exhibit a universal linear relationship despite the different FS topologies of the two compounds. It is thus reasonable to assume that the CDW order observed in the absence of an applied magnetic field is universally responsible for the reconstruction of the pseudogapped FS. 

Our calculation for Hg1201 points to a possible Lifshitz transition due to overlap of the electron pockets as the hole concentration approaches $p\sim0.105$ (Fig.~\ref{fig7}(a)), which would result in the formation of a hole pocket centered at the Brillouin zone center. This transition would significantly alter the electronic properties of Hg1201, with distinct signatures expected in transport experiments in high magnetic fields. 

\subsection{\label{sec:level2} Influence of disorder on the CDW order in Hg1201}

It has been suggested that disorder pins otherwise fluctuating CDW correlations and disrupts CDW coherence, rendering $\xi$ finite. \cite{Caplan15} This is supported by recent NMR work for YBCO that indicates static correlations pinned by native defects below $T_{CDW}$, although $\xi$ may indeed be set by the correlation length of the pure system. \cite{Wu15} 
Alternatively, CDW order might be long-ranged, static and unidirectional in the absence of disorder. \cite{Nie14} 

\begin{figure}%[!htb]
	\includegraphics[width=0.85\linewidth,angle=0,clip]{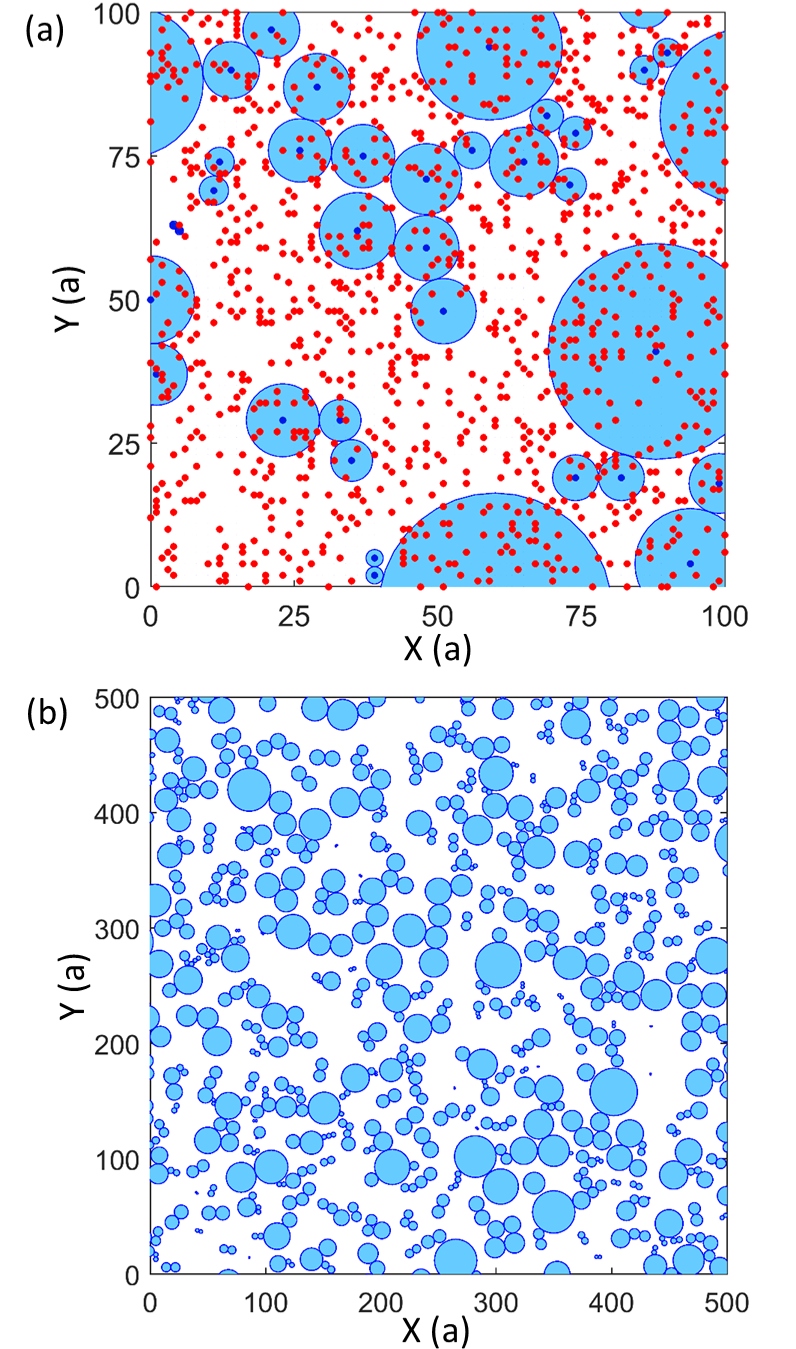}
	\caption{(color online) Representation of the two HgO$_{\delta}$ layers adjacent to a CuO$_2$ layer in Hg1201. (a) Red and blue dots represent randomly distributed i-O atoms with a density that corresponds to $p = 0.083$ (ref. \cite{Abakumov98,Antipov98,Putilin00}), the doping level at which the CDW correlation length is the largest (Fig. 6). The red dots indicate unit cells with one i-O, either in the top or the bottom HgO$_\delta$ layer. The dark blue dots indicate unit cells with two i-O, one in each of the two HgO$_\delta$ layers. In the model considered here (see also Appendix C), CDW patches (light blue circles) nucleate in these relatively rare unit cells and extend to the closest doubly-occupied site. This results in a CDW correlation length of $\xi/a \approx 7.7$, in good agreement with the experimental value $\xi/a = 8.0(7)$ at this doping level. With increasing doping, the calculated correlation length decreases and follows the experimental values, as shown in Fig.~\ref{SumAll}(b). Displayed area: $(100\,a)^2$. (b) Same as in (a), but on a larger scale (displayed area: $(500\,a)^2$) for better visualization of the CDW patch distribution.}\label{fig8}
\end{figure}

Although we are not able to experimentally distinguish between these scenarios, we can gain insight from considering the fact that a major source of disorder in Hg1201 are the interstitial oxygen (i-O) atoms in the HgO$_\delta$ layers. 
Under the assumption that the i-O atoms are randomly distributed, we have simulated the effects of these i-O atoms on the CDW correlation length (Appendix \ref{CLC}) and find that the average size of ordered patches pinned (in the form of halos around the pinning centers) or limited by i-O is very small, $\xi/a \approx 2.6$ at $p=0.083$, whereas the experimental value around this doping is $\xi/a \approx 7-8$ at low temperatures ($\xi/a \approx 4-5$ close to $T_{CDW}$). This suggests that individual i-O atoms have a rather weak effect. However, the local interactions should be more significantly altered by unit cells with two i-O atoms (one in each of the two HgO$_\delta$ layers adjacent to a CuO$_2$ layer). Indeed, in this case we find that the simulated correlation length agrees rather well with the low-temperature data (Fig.~\ref{fig8}). In this picture, the correlation length increases with doping, but the phase decoherence among neighboring CDW patches naturally limits their extend at doping levels above $p\approx 0.09$ (see Fig. \ref{SumAll}(a)). 
The increasing density of unit cells that host two i-O can explain why the CDW correlation length in Hg1201 never reaches the values observed for YBCO and why CDW order vanishes at lower doping. 
In YBCO, various types of oxygen-dopant defects are observed,\cite{Wu16} yet the proposed scenario involving i-O pairs is unique to Hg1201.
Hg1201 exhibits the highest optimal $T_c$ of all single-layer cuprates. \cite{Eisaki04} 
It is an intriguing possibility that, for Hg1201, $T_c$ at optimal doping is so high in part because the disorder potential of i-O pairs disrupts CDW order that would otherwise compete with SC order up to or even beyond optimal doping, as in YBCO or in single-layer electron-doped Nd$_{2-x}$Ce$_{x}$CuO$_{4+\delta}$. \cite{Neto16}

Disorder and electronic inhomogeneity are inherent to the cuprates. \cite{Phillips03,Eisaki04} NMR measurements indicate that, similar to other cuprates, Hg1201 exhibits a considerable local electric-field gradient distribution.\cite{Rybicki09} 
As a result, volume-averaging probes (e.g., RXS, XRD, optical spectroscopy and charge transport) integrate contributions from regions with different local electronic environments. 
This averaging might mask a possible inherent electronic instability at a distinct doping level where the CDW phenomenon is most robust ($p \approx 0.08$ in the case of Hg1201). 
On the other hand, if the CDW properties evolve monotonically as a function of doping, we expect the inhomogeneity to have little effect (as the averaged properties are effectively represented by the median value).

\subsection{\label{sec:level2} Nature of the atomic displacements associated with the CDW order in Hg1201}

We can furthermore make some inferences about the nature of the atomic displacements associated with the CDW order. In YBCO, the primary displacements are shear displacements in the $c$-direction.\cite{Forgan15} Because this material contains a bilayer, these displacements have a first-order effect on local carrier energy and doping, and therefore couple directly to the Fermi surface reconstruction. \cite{Briffa16} In Hg1201, the single CuO$_2$ plane per primitive cell is a mirror plane, and $c$-axis atomic displacements will only affect carrier energies to second order, and so will have a negligible coupling to the Fermi surface reconstruction. A similar argument also holds for basal-plane displacements transverse to the CDW wavevector. We therefore deduce that in Hg1201, the CDW in the CuO$_2$ planes must have a longitudinally-polarized component in order to couple to the Fermi-surface reconstruction.  

\begin{figure}%[!htb]
	\includegraphics[width=0.9\linewidth,angle=0,clip]{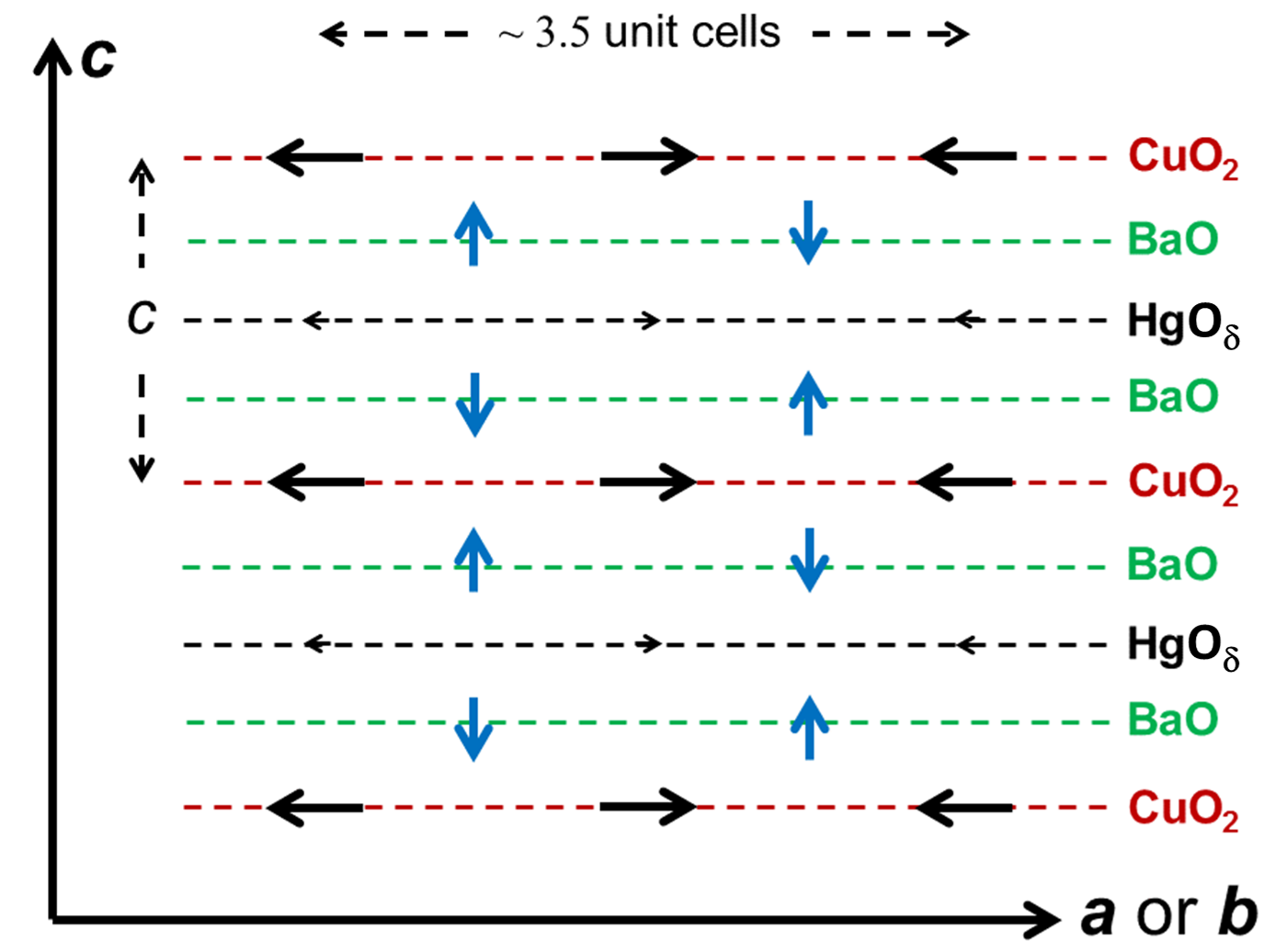}
	\caption{(color online) Illustration of the average atomic displacements proposed for the CDW order in Hg1201. The red, green and black dashed lines correspond to the CuO$_2$, BaO and HgO$_\delta$ layers, respectively. Black (blue) arrows represent the longitudinal (transverse) atomic displacements. Small and large displacements are indicated by the length of the arrows (not to scale). See the text for further details.
	}\label{fig9}  
\end{figure}

Figure \ref{fig9} contains a sketch of the atomic displacements required by group theory for a mode with a single incommensurate wavevector. The biaxial case is the superposition of two such patterns at right angles. Longitudinal displacements in a CuO$_2$ plane necessarily imply $c$-axis displacements in the BaO layers which are mirror-symmetric about the CuO$_2$ plane.  By symmetry, there will be no $c$-axis displacements in the HgO$_\delta$ layer, although there will be basal-plane displacements.

We have not measured sufficiently many CDW satellites to quantify the displacements represented in Fig. \ref{fig9}, which arise from physical and group theory arguments. However, we can show that this picture is consistent with the results obtained so far. First, we note that the contribution of an atomic displacement to the amplitude that determines the intensity of a CDW satellite is proportional to the scalar product of the scattering vector {\bf Q} and the displacement \cite{Forgan15}. Hence, the intensity at (3-$q_{CDW}$, 0, 1.25) primarily depends on basal-plane displacements, whereas the (4-$q_{CDW}$, 0, 6) satellite weights basal plane to $c$-axis amplitudes in the ratio $\sim1.5:1$. 
If out-of-plane displacements of Ba atoms dominated the structure factor, the $L=6$ satellite peak would be expected to be more than an order of magnitude stronger in intensity than the satellite at $L=1.25$, mainly due to a larger $L$ value. 
Since the intensities of the two satellites are comparable, this strongly suggests that the basal-plane displacements in the CuO$_2$ layer are the major ones involved in the CDW order. 
A full structural refinement, using a more relaxed geometry to allow many CDW satellites to be observed, would allow this picture to be confirmed.

For a single CDW mode, the Cu-O bond parallel to {\bf q} is modulated in length by the displacements, whereas the equivalent bond perpendicular to {\bf q} is not. This can be considered as an indication of a contribution from the $d$-density wave state originally proposed for YBCO.\cite{Sachdev13} It has been suggested that the breaking of fourfold symmetry around the planar copper atoms is a common aspect of the FS reconstruction in the cuprates. \cite{Ramshaw17} We see no evidence for a macroscopic symmetry breaking of this form in Hg1201. For a single CDW mode, the local fourfold symmetry around a copper atom is indeed broken. However, the superposition of two equivalent CDWs at right angles (the bidirectional case discussed earlier in this paper) would restore this symmetry globally.

%%%%%%%%%%%%%%%         SUMMARY       %%%%%%%%%%%%
	\section{\label{sec:level1} SUMMARY}
We have used synchrotron x-ray radiation to study the CDW order in simple tetragonal Hg1201. Our resonant x-ray scattering and hard x-ray diffraction measurements provide insight into this phenomenon and its connection with charge transport in the cuprates. The structural simplicity of Hg1201 enabled us to establish a direct link between properties of the CDW order and the Fermi-surface-reconstructed state. The consideration of our results along with available quantum-oscillation data allowed us to simulate the size of the Fermi pocket in Hg1201 and its evolution with doping. Although the observed doping and temperature dependences of the charge correlations in Hg1201 are similar to YBCO, the CDW order is more robust in the latter compound. 
Initially, the characteristic charge-order temperature and correlation length universally increase with increasing carrier concentration. In YBCO, the correlation length reaches a maximum of about 20 lattice units and CDW order is observed up to about optimal doping. In contrast, in Hg1201 the maximum is only about 8 lattice units and CDW order disappears well before optimal doping is reached, once the correlation length is comparable to the CDW modulation period. For Hg1201, the characteristic onset temperature of the CDW phenomenon as determined with x rays coincides with the characteristic temperature identified in prior transient reflectivity work. Consideration of the disorder induced by the interstitial oxygen atoms led us to propose that the CDW correlation length in Hg1201 is limited by pairs of such dopants within the same unit cell. Finally, our analysis of CDW satellite peaks observed via x-ray diffraction indicates that the dominant atomic displacements associated with the CDW order are longitudinally-polarized displacements within the CuO$_2$ plane.

%%%%%%% ACKNOWLEDGEMENTS  %%%%%%%%%%%%%%%%

\section{\label{sec:level1} ACKNOWLEDGMENTS}

We thank A. V. Chubukov, M.-H. Julien, Yuan Li, E. H. da Silva Neto, and C. Proust for valuable comments on the manuscript, and acknowledge the contributions of M. K. Chan, C. J. Dorow and M. J. Veit to the sample preparation. The work at the University of Minnesota (crystal growth and characterization, x-ray measurements and data analysis) was funded by the Department of Energy through the University of Minnesota Center for Quantum Materials, under DE-SC-0016371 and DE-SC-0006858. Research performed at the Canadian Light Source, was supported by CFI, NSERC, the University of Saskatchewan, the Government of Saskatchewan, Western Economic Diversification Canada, NRC, and CIHR. Work at HASYLAB and TUW was supported by FWF project P27980-N36. Work with the Birmingham 17 T cryomagnet was supported by the U.K. EPSRC, grant number  EP/J016977/1.  EMF was supported by the Leverhulme Foundation. The data were analyzed using SPECPLOT and SPCFIT packages; \url{http://webusers.spa.umn.edu/~yu/index.html}

\appendix

\section{Search for CDW order in nearly optimally doped Hg1201.} \label{OPD}
CDW order has been reported from (micro) x-ray diffraction measurements for a nearly-optimally-doped sample of Hg1201 ($T_c = 95$ K), and it has been suggested that the CDW order competes with i-O order. \cite{Campi15} The doping level of this sample is well above the hole concentration up to which we detected CDW order via resonant x-ray diffraction, a very sensitive probe of charge order. The reported CDW correlation length exceeds 15 lattice constants at the onset temperature, and hence is much larger than our result (Fig. 6a), and the reported wave vector of $q_{CDW} = 0.23$ r.l.u. is considerably lower than $q_{CDW} \approx 0.25$ r.l.u., the value predicted from linear extrapolation of our data in Fig.~\ref{SumAll}(b) to optimal doping. Moreover, the reported charge order sets in at 240 K, approximately 50 K above $T^*\approx 190$ K near optimal doping. \cite{Barisic15} 
These observations suggest that origin of the reported feature in ref. \cite{Campi15} is distinct from the CDW order reported here and for other cuprates.         
One possible explanation is that the observed peak originates from a spurious secondary phase; the data in Fig. 2a of ref.~\cite{Campi15} clearly show a secondary phase with larger lattice parameters, rotated by approximately 45 degrees with respect to the lattice of Hg1201. Two-phase coexistence would explain the apparent competition between the i-O order, a characteristic of Hg1201 near optimal doping. Furthermore, the different lattice constants (and thermal expansion coefficients) of the two phases would explain the unusual temperature dependence of the feature observed in ref. \cite{Campi15}

\section{Calculation of reconstructed Fermi surface} \label{FSC}

The tight-binding FS calculation (Fig.~\ref{fig7}) was performed for hole doping levels $0.07 \le p \le 0.11$. The hopping parameters that provide the best fit to dispersions from first-principles calculations were adopted from ref. \cite{Das12}: $(t, t', t'', t''') = (0.46,  0.105, 0.08, 0.02)$ eV. The doping dependence of the FS was simulated by adjusting the chemical potential so that the unreconstructed FS satisfies Luttinger's sum rule and contains $1 + p$ holes, consistent with photoemission spectroscopy data. \cite{Vishik14} Hence, $1+p=2A_{FS}/A_{BZ}$, where $A_{FS}$ ($A_{BZ}$) is the area of the Fermi surface (Brillouin zone). In order to simulate the reconstructed FS, following ref. \cite{Harrison11}, we constructed a 4 by 4 Hamiltonian and diagonalized it numerically at each point of reciprocal space, with a typical mesh of $10^6$ points. Having set the band dispersion and wave vector to the experimentally determined values, the only adjustable parameter left for the calculations is the CDW potential $\Delta_{CDW}$. We fixed the CDW potential to $\Delta_{CDW}=50$ meV so that at $p=0.09$ the size of the electron pocket that results from the FS reconstruction by the wave vector $q_{CDW} =0.279(5)$ r.l.u. yields the value of 3.06 \% of the Brillouin zone, consistent with experiment. \cite{Barisic13a} The CDW potential $\Delta_{CDW}$ was then kept constant for the calculation of the doping dependence, and the chemical potential was varied. The doping dependence of $q_{CDW}$ employed in the calculations was obtained from the linear fit to the data in Fig.~\ref{SumAll}(b). After diagonalizing the Hamiltonian, four bands were found to cross the Fermi level, but only the band that corresponds to the electron pocket is displayed in Fig.~\ref{fig7}; in order to obtain only one band crossing the Fermi level (the electron pocket), one would have to start with a pseudogapped FS. The resulting FS, plotted in the reduced Brillouin zone in Fig.~\ref{fig7}, is expected to undergo a Lifshitz transition from electron-like to hole-like Fermi pocket at $p\approx10.5\%$. This result could in principle be verified experimentally by performing high-field/low-temperature transport measurements.

\begin{figure*}[!]
	\includegraphics[width=1\linewidth,angle=0,clip]{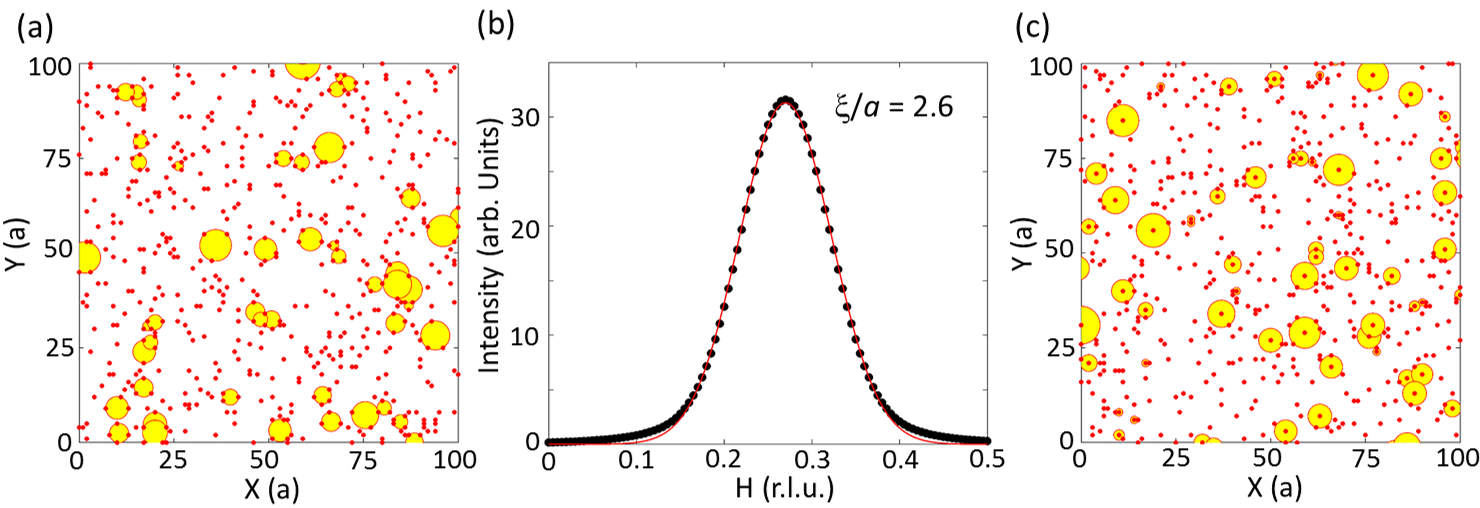}
	\caption{(color online) (a) CDW patches (yellow circles) in the CuO$_2$ planes formed in regions between i-O atoms (red dots) in the adjacent HgO$_\delta$ layers. Displayed area: $(100a)^2$. The i-O density corresponds to $p = 0.083$. \cite{Abakumov98,Antipov98,Putilin00} (b) CDW peak, calculated from the distribution of charge-order patches in (a); the correlation length $\xi/a = 2.6$  is obtained from a Gaussian fit. (c) Same as in (a), but CDW patches develop around i-O and extend to nearest-neighbor i-O. In both (a) and (c), the density of the charge-ordered patches has been decreased for clarity.}\label{fig10}
\end{figure*}

\section{Modeling of the CDW correlation length}\label{CLC}

Compared to YBCO, the CDW order in Hg1201 exhibits a rather small correlation length and vanishes well below optimal doping.
In search for a possible explanation for this observation, we consider the disorder that originates from the doped interstitial oxygen atoms (i-O) in the HgO$_\delta$ layers of Hg1201. 
We simulate the distribution of these i-O atoms and consider the resultant disorder length scales for three simple scenarios. 
We estimate the i-O density as a function of $T_c$ from ref. \cite{Putilin00,Abakumov98,Antipov98}
 
In the first scenario, we assume that CDW patches in the CuO$_2$ planes form within regions defined by the i-O and that the size of the CDW patches is limited by the neighboring i-O (Fig.~\ref{fig10}(a)). We randomly create circular CDW patches between the i-O, maximize their size, and numerically determine the probability $P(r)$ of finding a CDW patch with radius $r$. The scattering intensity from the CDW patches is given by
\begin{equation}\label{key}
I(Q) \propto \sum_{r}P(r)\abs{\int_{-r}^{r}dx\sqrt{r^2-x^2} Ae^{(iq_{CDW}\cdot x)}e^{-iQ\cdot x}}^2
\end{equation} 
\newline
where $A$ is the amplitude, $q_{CDW}$ is the magnitude of the CDW wave vector, and $Q$ is the magnitude of the momentum transfer. 
The calculations is performed on a 2000 x 2000 lattice of atoms for the doping levels $p = 0.063$, 0.083 0.091, 0.107 0.12 and 0.142.  
In the simulations, we assume that $A$ is independent of the radius. 
The size distribution of CDW patches results in a Gaussian line shape of the CDW peak (Fig.~\ref{fig10}(b)). 
The correlation length is obtained by fitting a Gaussian to the calculated CDW peak, in the same manner as the x-ray data. 
In this scenario, the calculated correlation lengths are significantly smaller than those obtained from experiment.  
For example, we obtain $\xi/a = 2.6$ for $p = 0.083$, whereas the experimental value at this doping level is $\xi/a = 8.0(7)$. 
This large discrepancy indicates that it is unlikely that CDW correlations are destroyed by disorder associated with individual i-O.
In the second scenario, we assume that the CDW patches develop around i-O and extend to the nearby i-O (Fig.~\ref{fig10}(c)).
The resultant correlation lengths are comparable to the first scenario.

In the third scenario, we consider that the local structural and electronic environment may be significantly different from the typical situation (with zero or one i-O atom) when a CuO$_2$ plaquette is affected by two i-O, one in each of the two adjacent Hg-O layers. Such relatively rare unit cells with two i-O atoms may act as nucleation centers of CDW order and/or as strong pinning sites that lead to the destruction of CDW coherence. 
In this case, CDW patches extend from unit cells with two i-O atoms to neighboring CDW patches with two i-O atoms, as shown in Fig.~\ref{fig8}. The phases of these CDW patches are uncorrelated, as the patches are pinned to different nucleation centers. This third scenario yields correlation lengths very close to the measured low-temperature values (Fig.~\ref{SumAll}(a)). A similar result is obtained by assuming that CDW patches are bound rather than pinned by such unit cells. Therefore, we cannot distinguish whether these sites assist the formation of CDW, or simply pin the CDW. We note that this third scenario is rather unique to Hg1201. For YBCO and other cuprates, the CDW correlations may be affected by other types of disorder.

\end{document}